\documentstyle[sprocl,epsf]{article}

\def\Journal#1#2#3#4{{#1} {\bf #2}, #3 (#4)}

\def\NPB{{\em Nucl. Phys.} B}
\def\PLB{{\em Phys. Lett.}  B}

\def\PRD{{\em Phys. Rev.} D}

\newcommand{\la}[1]{\label{#1}}
\newcommand{\be}{\begin{equation}}
\newcommand{\ee}{\end{equation}}
\newcommand{\ba}{\begin{eqnarray}}
\newcommand{\ea}{\end{eqnarray}}
\newcommand{\bi}{\begin{itemize}}
\newcommand{\ei}{\end{itemize}}

\newcommand{\tr}{{\rm Tr\,}}

\newcommand{\fr}[2]{{\frac{#1}{#2}}}

\newcommand{\lambdamsbar}{\Lambda_{\overline{\rm MS}}}%lambda_\msbar didnt work
\newcommand{\dr}{{4D\to3D}}
\newcommand{\eq}{Eq.\,}

\def\lsi{\raise0.3ex\hbox{$<$\kern-0.75em\raise-1.1ex\hbox{$\sim$}}}
\def\gsi{\raise0.3ex\hbox{$>$\kern-0.75em\raise-1.1ex\hbox{$\sim$}}}
\newcommand{\lsim}{\mathop{\lsi}}
\newcommand{\gsim}{\mathop{\gsi}}

\begin{document}

\title{DIMENSIONAL REDUCTION AND HOT QCD:
CALCULATING THE DEBYE MASS NON-PERTURBATIVELY\footnote{
Talk given at the conference ``Strong and Electroweak Matter '97'',
21--25 May 1997, Eger, Hungary.}}

\author{K. RUMMUKAINEN}

\address{Universit\"at Bielefeld, Fakult\"at f\"ur Physik, Postfach 100131, D-33501 Bielefeld,
Germany}


\vspace*{-1.2cm} \hfill BI-TP 97/27

\vspace*{0.3cm}

\maketitle\abstracts{
We study the phase diagram of the 3-dimensional
SU(2) + adjoint Higgs theory, and investigate to what extent
it can be used as an effective theory of the 4-dimensional
high-$T$ SU(2) QCD\@.  The relation between the parameters in
4 and 3 dimensions is obtained through dimensional reduction.
The high-$T$ (deconfined) QCD phase corresponds
to the {\em symmetric} phase of the 3-D Higgs theory.  In the
relevant parameter region the symmetric phase is not stable,
but the metastability is strong enough to make precise measurements
possible.  In particular, we measure the Debye mass using a
gauge invariant operator.}

Numerical lattice Monte Carlo studies of finite temperature QCD are
very costly due to the difficulty of the fermionic action simulations.
Nevertheless, the effects of fermions are essential at high $T$ and
cannot be ignored.\cite{Ukawa} This problem can be partly overcome
with the {\em dimensional reduction}
(DR)~\cite{Ginsparg}$^{\!-\,}$\cite{su2paper}: it provides a method for
obtaining an {\em effective 3D bosonic theory} for the full 4D finite
$T$ QCD.  The 3D theory can be derived perturbatively without the
infrared problems associated with the standard high-$T$ perturbative
analysis.  It retains the essential infrared physics of the original
theory, and since it is bosonic, it can be studied very economically
with lattice Monte Carlo simulations.  Recently it has been
successfully applied to the Electroweak phase transition.\cite{ewreview}

In this paper we apply DR to SU(2) QCD, and in particular we determine
the ${\cal O}(g^2T)$ correction to the Debye screening mass in the
deconfined phase.  This is not computable in perturbation
theory.\cite{ay}$^{\!,\,}$\cite{Rebhan} 
%The action of SU(2) QCD with $N_f$ fermions in 4D is
%\be
%S_4=\int_0^\beta\!d\tau\int\! d^3x\biggl\{\fr14 F_{\mu\nu}^aF_{\mu\nu}^a+
%\sum_i\bar\psi_i[\gamma_\mu D_\mu(A)+m_i]\psi_i\biggr\}\,.
%\ee

The dimensionally reduced effective theory of the 4D SU(2) QCD with $N_f$
fermions is a 3D SU(2) + adjoint Higgs theory:
\be
S_3 = \int\! d^3x \biggl\{
\fr14  F_{ij}^aF_{ij}^a
+ \tr [D_i,A_0][D_i,A_0]
+ {m_D^2} \tr A_0^2 + {\lambda_A}(\tr A_0^2)^2 \biggr\}
\la{3daction}
\ee
The Higgs field $A_0$ is a remnant of the temporal gauge fields and
belongs to the adjoint representation of SU(2).  The dimensional
reduction is performed at 2-loop level for the couplings $m_D^2$ and
$\lambda_A$, and at 1-loop level for the 3D gauge coupling $g_3^2$
(for details, see~\cite{su2paper}).  The couplings are dimensionful,
and it is convenient to re-express them as a one dimensionful scale
and two dimensionless parameters $x$, $y$:
\be
  g_3^2, ~~~ y \equiv \frac{m_D^2}{g_3^4}, ~~~ 
  x \equiv \frac{\lambda_A}{g_3^2} \,.
\ee
For $N_f=0$, the 3D couplings
are related to the temperature $T$ (and the 4D gauge coupling
$g^2(\mu)$, which is evaluated at the optimized scale~\cite{su2paper}
$\mu \approx 2\pi T$) by
\ba
g_3^2 &=& g^2 T = \fr{24\pi^2 T}{22 \log(6.742 T/\lambdamsbar)} \la{g3}\\
x &=& \fr{\lambda_A}{g_3^2} = \fr{10}{22}
	\fr1{\log(5.371T/\lambdamsbar)}   \la{x} \\
y &=& \fr{2}{9\pi^2 x} \left( 1 +\fr98 x\right) \la{y}
\ea
The presence of fermions only modifies the numerical factors in the
above equations.  Most of our results are from $N_f=0$ -case, but
measuring the same observables for $N_f > 0$ would be just equally
straightforward.

%%%%%%%%%%%%%%%%%%%%%%%%%%%%%%%%%% FIGURE
%\begin{figure}[tb]

%\vspace*{-0.5cm}

%\epsfysize=13cm
%\centerline{\epsffile{h_a0_mass.eps}}

%\vspace*{-5cm}

%\caption[a]{The masses of the $A_0^2$ (solid) and $h$ (opaque) operators.}
%\la{fig:h_a0}
%\end{figure}
%%%%%%%%%%%%%%%%%%%%%%%%%%%%%%%%%%%%%

A very important feature of the action given in Eq.~\ref{3daction} is that
it is {\em
superrenormalisable\,}: it has only linear 1-loop and logarithmic
2-loop divergences.  By calculating the divergent counterterms both
in the continuum and on the lattice we obtain a definite set of
equations
linking the 3D continuum coupling constants ($g_3^2,\, x,\, y$) 
to the corresponding lattice couplings and the lattice spacing.  
In particular, the dimensionless coupling $\beta_G$, which multiplies
the lattice gauge action (plaquette term), is related to the
continuum gauge coupling $g_3^2$ and the lattice spacing $a$ by
\be
 \beta_G = \fr{4}{g_3^2\,a}\,.  \la{betag}
\ee
A detailed discussion of the lattice action and the continuum
$\leftrightarrow$ lattice connection is given in ref.~\cite{su2paper}.

%%%%%%%%%%%%%%%%%%%%%%%%%%%%%%%%% FIGURE
\begin{figure}[tb]

\vspace*{-0.5cm}

\epsfysize=12.5cm
\centerline{\epsffile{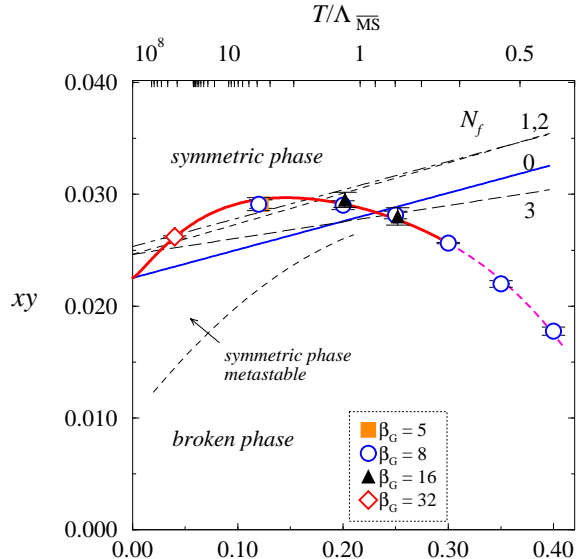}}

\vspace*{-5cm}

\caption[a]{The phase diagram of the 3D SU(2) + adjoint Higgs
theory.  The plot symbols with error bars are lattice results
for $y_c(x)$ (extrapolated to $V\to\infty$), and the thick
line is a 4th order polynomial fit to the data.
The dashed line
marks the region where the transition turns into a cross-over.
The straight lines are the $y_\dr(x)$ curves for the indicated
$N_f$ -values (for $N_f=0$, the line is given by \eq\ref{y}).
The top scale shows the values of
$T/\lambdamsbar$ corresponding to the values of $x$ along $y_\dr(x)$ 
for $N_f=0$ (\eq\ref{x}).}
\vspace*{-3mm}
\la{fig:phasediag}
\end{figure}
%%%%%%%%%%%%%%%%%%%%%%%%%%%%%%%%%%%%

The phase diagram of the SU(2) + adjoint Higgs theory is shown in
Fig.\,\ref{fig:phasediag}.  For convenience, it is plotted in the
($x,xy$)-plane; the third (dimensionful) coupling $g_3^2$ gives the
length scale.  The plot symbols indicate the location of the
transition given by the Monte Carlo simulations with different lattice
spacings ($\beta_G$, Eq.~\ref{betag}).  Each of the plot symbols
includes a $V\rightarrow\infty$ extrapolation using a series of
simulations at finite volumes.  The {\em critical curve\,}
$y=y_c(x)$ shown is a 4th order polynomial interpolation of the data.  Above
$y_c(x)$ the system is in the symmetric phase, below it in the broken
phase.  At small $x$, the transition is very strongly first order, but
becomes rapidly weaker when $x$ increases.  At $x\approx 0.3$--0.33
there is a critical point, after which only a cross-over remains.

The straight lines in Fig.~\ref{fig:phasediag} are the `dimensional
reduction lines' $y_\dr(x)$ for the number of fermion flavors
indicated.  The $N_f=0$ line is the pure gauge line given in
Eq.~\ref{y}.  The 3D theory is well defined on the entire
$(x,y)$-plane, but only along $y_\dr(x)$ the 3D theory can describe
the physics of the 4D SU(2) gauge theory.  Along this line $x$ is
related to the temperature as shown on the top axis of the plot (for
SU(2), $\lambdamsbar \approx 1.2 T_c$~\cite{fingberg}).

Note that in the physically relevant region $T \gg T_c \sim
\lambdamsbar$ the physical $y_\dr$ line is in the {\em broken
phase\,}.\footnote{ This is in contrast to the earlier result by the
Bielefeld group,\cite{bielefeld} which observed the symmetric phase to
be absolutely stable.  The difference is due to the increased accuracy
in the reduction $\dr$ used here.}\, However, the broken phase cannot
describe 4D physics since the perturbative analysis used to derive
Eq.~\ref{3daction} is not valid there.  Nevertheless, the simulations
along $y_\dr(x)$ can still be performed in the symmetric phase: we
utilize the fact that due to the strong 1st order nature of the
transition at small $x$ the symmetric phase is strongly {\em
metastable\,}.  The region of metastability is shown with the dashed
line in Fig.~\ref{fig:phasediag}.  Excepting microscopic lattice
volumes, the system remains in the symmetric phase for the duration of
any realistic Monte Carlo simulation.

%%%%%%%%%%%%%%%%%%%%%%%%%%%%%%%%% FIGURE
\begin{figure}[tb]

\vspace*{-0.5cm}

\epsfysize=12.5cm
\centerline{\epsffile{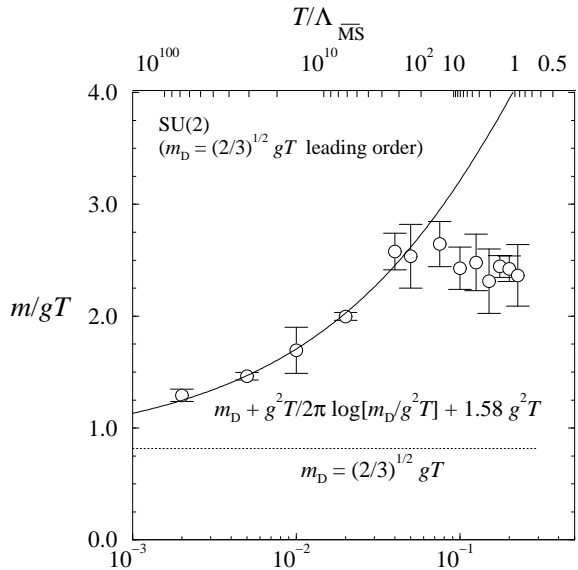}}

\vspace*{-5cm}

\caption[a]{The Debye mass from operator $h$, in units of
$gT$.  The line is the 1-parameter fit to Eq.~\ref{mdnonp}.\la{fig:debyemass}}
\vspace*{-3mm}
\end{figure}
%%%%%%%%%%%%%%%%%%%%%%%%%%%%%%%%%%%%

We measure the Debye screening mass along the $y_\dr(x)$ -line
with the operator~\cite{ay}$^{\!,\,}$\footnote{
This operator was also used by Hart et al.\cite{hart}; however,
not in the metastable symmetric phase corresponding to the $y_\dr(x)$
-lines.} 
\be
   h_i = \varepsilon_{ijk} \tr A_0 F_{jk}\,.
\ee
This operator is the lowest dimensional gauge invariant operator
containing only one insertion of $A_0$.  

When the ${\cal O}(g^2T)$ corrections to the Debye mass are included,
\be
   m_D = m_{D,0} + \fr{g^2 T}{2\pi} \log \fr{m_{D,0}}{g^2 T} +
C_2\, g^2 T  + {\cal O}(g^3 T)\,,  \la{mdnonp}
\ee
where $m_{D,0} = \sqrt{2/3}gT$ is the leading order contribution and the
coefficient in front of the log-term is a result of a perturbative
computation of the pole $A_0$ mass.\cite{Rebhan} The coefficient
{$C_2$} is not computable perturbatively.

To improve the signal we use recursive smearing and blocking
of both the gauge fields and the Higgs field.  In order to observe the 
true asymptotic behavior we use very small values of $x$, which
correspond to extremely large $T$ -- up to $10^{100}\times T_c$.

In Fig.~\ref{fig:debyemass} we show the data in units of $gT$
along the $y=y_\dr(x)$ -line (for $N_f=0$).  The temperature
in physical units is shown in the top scale of the figure.
When $x\lsim 0.05$ ($T \gsim 100 \lambdamsbar \sim 100 T_c$)
the asymptotic function \eq\ref{mdnonp} fits the data well, with
the result
\be
  C_2 = 1.58 \pm 0.18\,.
\ee
From Fig.~\ref{fig:debyemass} we observe that at all realistic
temperatures, the corrections to the leading order result $m_{D,0}$
are very large.  $m_{D,0}$ dominates only when $x\lsim 0.01$,
corresponding to temperatures $T \gsim 10^{20} \lambdamsbar$!
Interestingly, when $T\lsim 100 T_c$, the mass remains approximately
constant, $m_D \sim 3\,m_{D,0}$.  Large values for $m_D$ have also
been observed in 4D SU(2) simulations, using gluon propagators in
Landau gauge.\cite{rank}

\end{document}